\documentclass[runningheads]{llncs}
\usepackage{hyperref}
\usepackage{scalerel}
\usepackage{tikz}
\usetikzlibrary{svg.path}
\definecolor{orcidlogocol}{HTML}{A6CE39}
\tikzset{
  orcidlogo/.pic={
    \fill[orcidlogocol] svg{M256,128c0,70.7-57.3,128-128,128C57.3,256,0,198.7,0,128C0,57.3,57.3,0,128,0C198.7,0,256,57.3,256,128z};
    \fill[white] svg{M86.3,186.2H70.9V79.1h15.4v48.4V186.2z}
    svg{M108.9,79.1h41.6c39.6,0,57,28.3,57,53.6c0,27.5-21.5,53.6-56.8,53.6h-41.8V79.1z M124.3,172.4h24.5c34.9,0,42.9-26.5,42.9-39.7c0-21.5-13.7-39.7-43.7-39.7h-23.7V172.4z}
    svg{M88.7,56.8c0,5.5-4.5,10.1-10.1,10.1c-5.6,0-10.1-4.6-10.1-10.1c0-5.6,4.5-10.1,10.1-10.1C84.2,46.7,88.7,51.3,88.7,56.8z};
  }
}
\newcommand\orcidicon[1]{\href{https://orcid.org/#1}{\mbox{\scalerel*{\begin{tikzpicture}[yscale=-1,transform shape]\pic{orcidlogo};\end{tikzpicture}}{|}}}}

\usepackage{graphicx}
\usepackage{multirow}
\usepackage{tabularx}
\usepackage[numbers,sort&compress]{natbib}
\usepackage{float}

\begin{document}
\title{ABCD Neurocognitive Prediction Challenge 2019: Predicting individual fluid intelligence scores from structural MRI using probabilistic segmentation and kernel ridge regression}
\author{
Agoston Mihalik\inst{1,2,*}\protect\orcidicon{0000-0002-4510-4933} \and
Mikael Brudfors\inst{1,3,*}\protect\orcidicon{0000-0002-2884-2336}\and 
Maria Robu\inst{1,4}\protect\orcidicon{0000-0003-0106-0542} \and
Fabio S. Ferreira\inst{1,2}\protect\orcidicon{0000-0002-0977-2539} \and
Hongxiang Lin\inst{1}\protect\orcidicon{0000-0001-6643-327X} \and
Anita Rau\inst{1,4}\protect\orcidicon{0000-0002-4759-2846} \and
Tong Wu\inst{1,2}\protect\orcidicon{0000-0002-7468-2249} \and
Stefano B. Blumberg\inst{1}\protect\orcidicon{0000-0002-7150-9918} \and
Baris Kanber\inst{5}\protect\orcidicon{0000-0003-2443-8800} \and
Maira Tariq\inst{1}\protect\orcidicon{0000-0002-2826-4046} \and
Maria Del Mar Estarellas Garcia\inst{1} \and
Cemre Zor\inst{1,2}\protect\orcidicon{0000-0002-6141-2610} \and
Daniil I. Nikitichev\inst{1,4}\protect\orcidicon{0000-0001-5877-9174} \and
Janaina Mourao-Miranda\inst{1,2,\#}\protect\orcidicon{0000-0002-3309-8441} \and
Neil P. Oxtoby\inst{1,\#}\protect\orcidicon{0000-0003-0203-3909}}

\authorrunning{A.~Mihalik \& M.~Brudfors, et al.}
\titlerunning{Predicting IQ from MRI using probabilistic segmentation \& ridge regression}
\institute{%
Centre for Medical Image Computing (CMIC), \newline{}Department of Computer Science \& \newline{}Department of Medical Physics and Biomedical Engineering, \and
Max Planck UCL Centre for Computational Psychiatry and Ageing Research, \and
The Wellcome Centre for Human Neuroimaging, \and
Wellcome/EPSRC Centre for Interventional and Surgical Sciences (WEISS), \and
Department of Clinical and Experimental Epilepsy, \newline{}Queen Square Institute of Neurology;\\ \vspace{3mm}
University College London, Gower Street, London, WC1E 6BT, United Kingdom \\ \vspace{3mm}
\textasteriskcentered\,\# These authors contributed equally to this work.}

\maketitle

\begin{abstract}
We applied several regression and deep learning methods to predict fluid intelligence scores from T1-weighted MRI scans as part of the ABCD Neurocognitive Prediction Challenge (ABCD-NP-Challenge) 2019. We used voxel intensities and probabilistic tissue-type labels derived from these as features to train the models. The best predictive performance (lowest mean-squared error) came from Kernel Ridge Regression (KRR; $\lambda=10$), which produced a mean-squared error of $69.7204$ on the validation set and $92.1298$ on the test set. This placed our group in the fifth position on the validation leader board and first place on the final (test) leader board.
\end{abstract}

\section{Introduction}\label{INTRO}

Establishing the neurobiological mechanisms underlying intelligence is a key area of research in Neuroscience \cite{Goriounova2019,Deary2010}. General intelligence at a young age is predictive of later educational achievement, occupational attainment, and job performance \cite{McCALL1977,Gottfredson1997,Deary2007,Johnson2006}. Moreover, intelligence in childhood or early adulthood is associated with health outcomes later in life as well as mortality \cite{Batty2007,Batty2008,Lam2017,Deary2007,Deary2013}. Thus, understanding the mechanisms of cognitive abilities in children potentially has important implications for society and can be used to enhance such abilities, for example through targeted interventions such as education and the management of environmental risk factors \cite{Gottfredson1997,Fors2018}.

Neuroimaging can play a key role in advancing our understanding of the neurobiological mechanisms of cognitive ability. Several brain-imaging studies have shown that total brain volume is the strongest brain imaging derived predictor of general intelligence \cite{MacLullich2002,McDaniel2005,Rushton2009} ($r\approx0.3-0.4$). To a somewhat lesser degree, regional cortical volume and thickness differences in the frontal, temporal, and parietal lobes have also been linked to intelligence
\cite{Andreasen1993,Narr2007,Karama2011,McDaniel2005,MacLullich2002}. Converging neuroimaging evidence led to the proposal of the parieto-frontal integration theory \cite{Jung2007} whereby a distributed network of brain regions is responsible for the individual variability in cognitive abilities. This theory is also supported by human lesion studies \cite{Glascher2009,Woolgar2010}.

The ABCD-NP Challenge 2019 asked the question \textit{``Can we predict fluid intelligence from T1-weighted MRI?''} We took an exploratory, data-driven approach to answering this question --- a hackathon organised by our local research centres: the UCL Centre for Medical Image Computing (CMIC) and Wellcome/EPSRC Centre for Interventional and Surgical Sciences (WEISS). Our centres aim to address key medical challenges facing 21st century society through world-leading research in medical imaging, medical image analysis, and computer-assisted interventions. Our expertise extends from feature extraction/generation through to image-based modelling \cite{Oxtoby2017}, machine learning \cite{Schrouff2013,Schrouff2018,Blumberg2018}, and beyond. The hackathon involved researchers across research groups in our centres, in addition to colleagues from the affiliated Wellcome Centre for Human Neuroimaging and the Department of Clinical and Experimental Epilepsy at UCL. The hackathon took place on an afternoon in February 2019, after which we followed up with regular progress meetings.

In this paper we report our findings for predicting fluid intelligence in 9/10-year-olds from T1-weighted MRI using machine learning regression and deep learning methods (convolutional neural networks --- CNNs). Our paper is structured as follows. The next section describes the challenge data and our methods. Section~\ref{RESULTS} presents our results, which we discuss in section~\ref{DISCUSSION} before concluding.

\section{Methods}\label{METHODS}
\subsection{Data}

The ABCD-NP Challenge data consists of pre-processed T1-weighted MRI scans and fluid intelligence scores for children aged {9--10 years}. The imaging protocol can be found in \cite{Casey2018a}. Pre-processing included skull-stripping, noise removal, correction for field inhomogeneities \cite{Hagler2018,Pfefferbaum2018}, and affine alignment to the SRI24 adult brain atlas \cite{Rohlfing2010}. SRI24 segmentations and corresponding volumes were also provided. 

The cohort was split into training ($N=3739$), validation ($N=415$), and test ($N=4515$) sets. The training and validation sets also include scores of fluid intelligence, which are measured in the ABCD Study using the NIH Toolbox Neurocognition battery \cite{Akshoomoff2013}. For the challenge, fluid intelligence was residualised to remove linear dependence upon brain volume, data collection site, age at baseline, sex at birth, race/ethnicity, highest parental education, parental income and parental marital status. 

\subsection{Features derived from the data}\label{sec:features}

We trained the models to predict fluid intelligence both from the provided T1-weighted images (voxel intensity) as well as voxel-wise feature maps generated from these images using a probabilistic segmentation approach. There are many different methods to extract various features from T1-weighted images \cite{Monte-Rubio2018}, such as tissue-type labels obtained from probabilistic segmentations. These segmentations can be constructed in a way to capture not only the relative tissue composition in a voxel, but also information about shape differences between individuals. This requires mapping each subject to a common template --- a fundamental technique of computational anatomy \cite{Ashburner2007}. Here we constructed such a template from all available T1-weighted MRI scans ($S = train + validation + test = 3,739 + 415 + 4,515 = 8,669$), which generated normalised (non-linearly aligned to a common mean) tissue segmentations for each subject.

\begin{figure*}[ht]
 {\centering
  \includegraphics[width=0.9\textwidth]{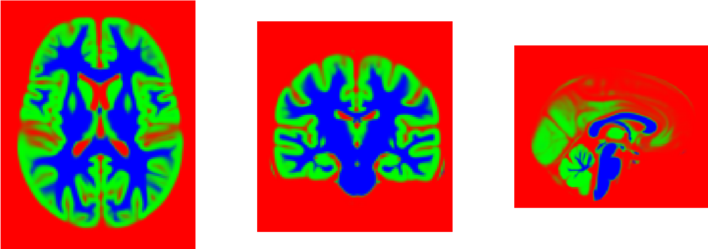} 
  \caption{Template generated from fitting the generative model to all of the subjects in the ABCD population ($S = 8,669$). Green corresponds to grey matter tissue, blue to white matter, and red to other.} 
  \label{fig:ProbSeg}
 }
\end{figure*}

We used a generative model \cite{Blaiotta2018} to probabilistically segment each T1-weighted MRI in the challenge data set into three tissue types: grey-matter, white-matter, other --- see Figure~\ref{fig:ProbSeg}. The in-house model\footnote{Available from \url{https://github.com/WCHN/segmentation-model}.} used here contains key improvements over the one in \cite{Blaiotta2018}: (1) we place a smoothing prior on the template; (2) we obtain better initial values by first working on histogram representations of the images; (3) we normalise over population image intensities in a principled way, within the model; (4) we place a prior on the proportions of each tissue, which is also learned during training. There are two types of normalised segmentations: non-modulated and modulated. Modulated segmentations include the relative shape change when aligning to the common template. 

Seven features per voxel were considered: T1-weighted intensity plus our six derived features corresponding to modulated and non-modulated probabilities for each tissue type. All images, including the feature maps from probabilistic segmentations, were spatially smoothed with a Gaussian kernel of 12mm FWHM \cite{Monte-Rubio2018} and masked to remove voxels outside of the brain.

\subsection{Predicting fluid intelligence: Machine Learning Regression}\label{predmodReg}

We explored several machine learning regression algorithms of varying complexity, including Multi-Kernel Learning (MKL) \cite{Rakotomamonjy2008}, Kernel Ridge Regression (KRR) \cite{Shawe-Taylor2004}, Gaussian Process Regression (GPR) \cite{Rasmussen2006} and Relevance Vector Machines \cite{Tipping2001}. The inputs to these models consisted of different concatenated combinations of our seven voxel-wise features described in section~\ref{sec:features}. Analyses were run in \emph{PRoNTo version 3} \cite{PRONTO,Schrouff2013}, a software toolbox of pattern recognition techniques for the analysis of neuroimaging data, as well as custom-written code.

In our preliminary analyses, we trained different combinations of regression algorithms and features using 5-fold cross-validation within the training set to select the best combination of algorithm and features as measured by lowest cross-validated mean-squared error (MSE). We then rtrained the best-performing model using the entire training set and used this trained model to generate predictions of fluid intelligence scores for the validation and test sets. Our best-performing model was KRR using all six voxel-wise derived features (tissue-type probabilities) concatenated into an input feature vector of length $\sim1.7$ million per individual. We set the regularization hyperparameter to $\lambda=10$ \cite{Shawe-Taylor2004}, which was optimised through 5-fold nested cross-validation within the training set in preliminary analyses.

We investigated robustness/stability of our KRR model using modified jackknife resampling (80/20 train/test split). Explicitly, we trained the model on a random subsample of $80\%$ of the training set and generated predictions for both the held out $20\%$ of the training set and the full validation set. We repeated this procedure 1000 times to generate confidence bounds on performance (MSE).

\subsection{Predicting fluid intelligence: Convolutional Neural Networks}\label{predmodCNN}

Separately, we explored the use of CNNs. The motivation was to incorporate spatial information that is not explicitly modelled as features. The CNNs were trained directly on the pre-processed T1-weighted images. Similarly to previous work that predicted brain age, Alzheimer's disease progression, or brain regions from MRI scans \cite{Sturmfels2018,Payan2015,Milletari2016} we applied various layers of 3D-convolutional kernels with filter size of 3x3x3 voxels on down-sampled images with dimensions of 61x61x61 voxels. We trained and validated multiple neural networks including those in \cite{Blumberg2018}. Our best performing CNN (lowest MSE) consisted of four convolutional layers and three fully-connected layers followed by dropout layers with a probability of $0.5$. The first six layers were activated by rectified linear units. The convolutional layers were followed by batch normalization and max-pooling operations. We used the Adam optimizer with an initial learning rate of $10^{–5}$ and a decay of $10^{–5}$, and we stopped training at the $100^{th}$ epoch. To evaluate the network, we randomly sampled 10 subsets of 1870 subjects from the training set, applied them to train 10 CNN models and evaluated performance on the validation set. We report MSE averaged over the ten passes.

\section{Results}\label{RESULTS}
Our best-performing regression model was KRR using our six derived voxel-wise features as input. This produced $\mathrm{MSE}= 69.72$ on the validation set. By comparison, our best-performing CNN achieved $\mathrm{MSE}= 70.82$ (average $\mathrm{MSE}= 73.40$) on the validation set. However, we observed that the better-performing CNNs on the training set did not generalise well to the validation set, possibly due to over-fitting, or mismatch between training and validation data (as suggested by the considerable difference in variances).

Table~\ref{Table-2} shows MSE and Pearson's correlation coefficient (mean $\pm$ std) for training and validation predictions from our top-performing methods. We found that KRR performed the best and we used this model to generate our challenge submission, which resulted in $\mathrm{MSE} = 92.13$ on the test set.

\begin{table}[ht]
\caption{Model performance for the training and validation sets.}\label{Table-2}
{\centering
\bgroup
\def\arraystretch{1.5}
\setlength\tabcolsep{3.5pt}
\begin{tabular}{|c|c|c|c|c|c|}
\hline
\multirow{2}{*}{\textbf{Method}} & \multicolumn{2}{c|}{\textbf{Training set}} & \multicolumn{2}{c|}{\textbf{Validation set}} & \textbf{Test set} \\ \cline{2-6}
    & \textbf{MSE} & \textbf{Correlation} & \textbf{MSE}  & \textbf{Correlation} & \textbf{MSE}   \\\hline
\textbf{KRR}    &     77.64       &        0.1427  &  \textbf{69.72}     &0.0311  & \textbf{92.13}       \\ \hline
\multirow{2}{*}{\textbf{CNN}}      &     $69.10 \pm 3.13$   &  $0.1913 \pm 0.0423$  & $73.40 \pm 2.13$     & $-0.0202 \pm 0.0296$  & N/A   \\ 
& (best: 64.39) & (best: 0.2542) & (best: 70.82) & (best: 0.0157) &\\
\hline
\end{tabular}
\egroup
}
\end{table}

\begin{figure*}[ht]
 \begin{center}
  \includegraphics[trim=0 4 0 11,clip,width=0.75\textwidth]{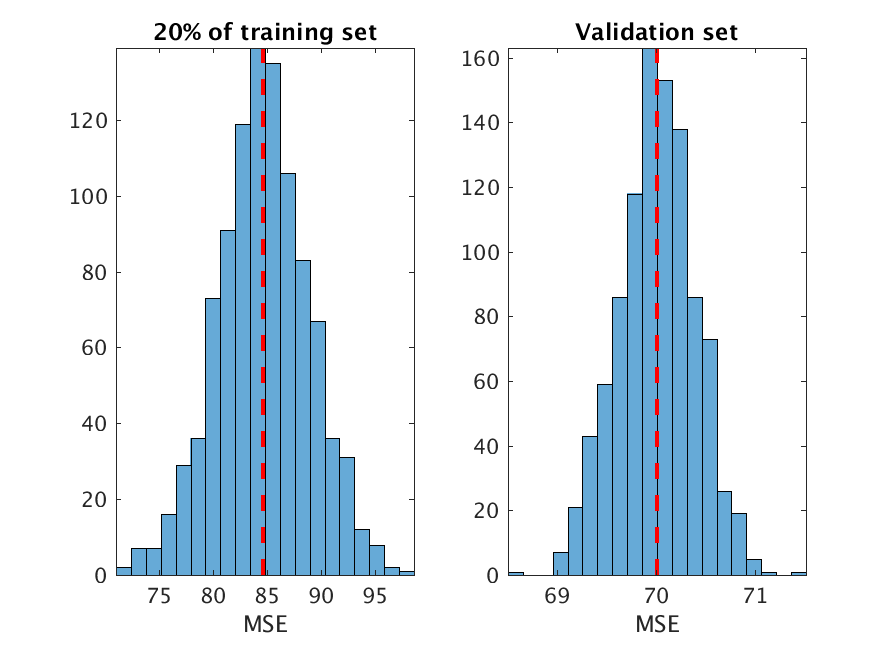} 
  \caption{Model robustness for KRR using modified jackknife re-sampling. MSE distributions for 1000 predictions: $20\%$ of the training set (left); full validation set (right); for models trained on the other $80\%$ of the training set. Red line indicates the mean.} 
  \label{fig:MSE}
 \end{center}
\end{figure*}

Figure~\ref{fig:MSE} shows model robustness results from modified jackknife resampling (1000 repetitions, 80/20 split). Confidence in our predicted MSE on the validation set is within $\pm1$ residual IQ point.

\begin{figure*}[ht]
 \begin{center}
  \includegraphics[width=0.85\textwidth]{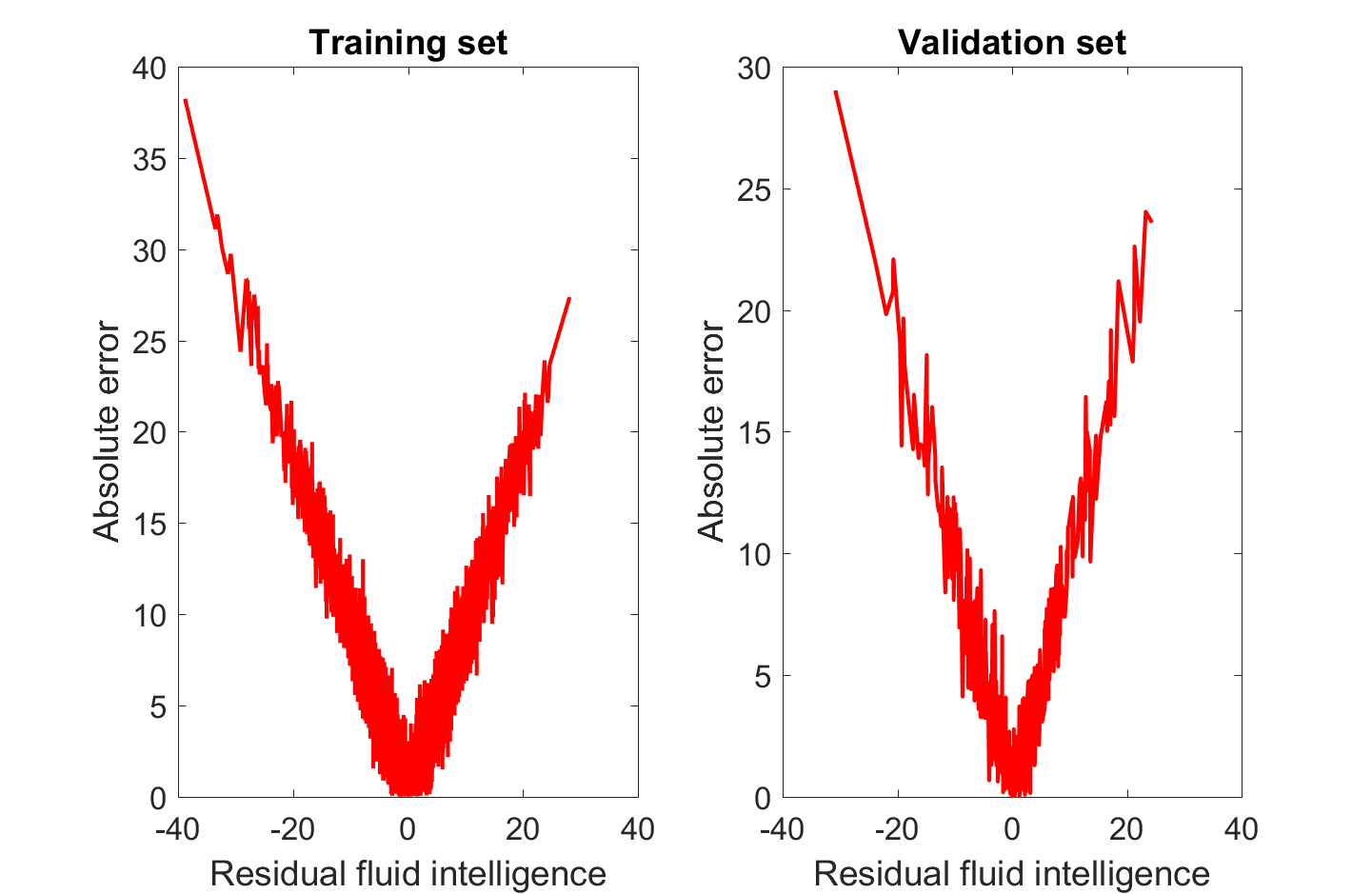}
  \caption{Absolute prediction errors against residual fluid intelligence scores for KRR (training set on the left, validation set on the right).}\vspace{-3mm} 
  \label{fig:error}
 \end{center}
\end{figure*}

Figure~\ref{fig:error} shows the residual fluid intelligence score prediction errors for KRR on the training and validation sets. The V shape of the curves shows that smaller residual scores have lower errors, since the model is predicting close to the mean value.

\newpage
\section{Discussion}\label{DISCUSSION}
We found that predicting residual fluid intelligence from structural MRI images is challenging. The correlation between predicted and actual intelligence scores was low for all methods we tested ($r\approx0.05-0.15$). This contrasts with previous studies for predicting (non-residualised) fluid intelligence, which have demonstrated that both total brain volume and regional cortical volume/thickness differences are relatively strong predictors ($r\approx0.3-0.4$) \cite{McDaniel2005,Rushton2009,MacLullich2002}.

The lower predictive performance we observed might be influenced by the residualisation, which prevents modelling of covariance between the residualisation factors and the image-based features. Moreover, there is evidence that including variables in the residualisation procedure that are correlated with the regression targets/labels is likely to remove important variability in the data leading to predictive models with low performance \cite{Rao2017}.

We note that previous studies used small sample sizes (of order 10--100), whilst the ABCD-NP Challenge dataset comprised a very large-scale dataset (of order 1000--10000). Whereas subject recruitment for small samples tend to be well controlled, resulting in homogeneous sample characteristics, large samples are more heterogeneous by nature, thus predictive models are more challenging to build. Accordingly, a recent study has demonstrated that the accuracy of classification results tends to be smaller for larger sample sizes \cite{Gael2018}.

Our image-based features were voxel-wise probabilistic tissue-type labels: grey matter, white matter, and other. Beyond tissue-type labels, there might be value in investigating other features generated by the generative segmentation model. One such interesting feature is scalar momentum \cite{Monte-Rubio2018}, which has been shown to be predictive for a range of different problems \cite{Monte-Rubio2018}.

Finally, we mention a possible limitation of our approach. In recent years, it has become standard practice to create study-specific group templates in neuroimaging \cite{Ashburner2007}, especially in Voxel-Based Morphometry (VBM) analyses. Originally, this approach was proposed for group analysis using mass univariate statistics (e.g., statistical parametric mapping). However, one needs to exercise caution when applying such an approach in machine learning, as it might lead to slightly optimistic predictions by creating dependence across the overall dataset. In order to avoid this potential issue, one would need to create templates based only on the training set. This might be computationally challenging when cross-validation strategies are used. To the best of our knowledge, no studies have investigated whether study-specific templates indeed result in slightly inflated predictions, and it remains an interesting question for future work.

\section{Conclusion}

Our paper presents the winning method for the ABCD Neurocognitive Prediction Challenge 2019. We found that kernel ridge regression outperformed more complex models, such as convolutional neural networks, when predicting residual fluid intelligence scores for the challenge dataset using our custom tissue-type features derived from the preprocessed T1-weighted MRI. The correlation between the predicted and actual scores is very low ($r=0.03$ for the KRR on the validation set), implying that the association between structural images and residualised fluid intelligence scores is low. It may be that structural images contain very little information on residualised fluid intelligence, but further study is warranted.

\bibliography{ABCD_KRR}


\end{document}